\documentclass[letterpaper]{article}
\pdfoutput=1
\usepackage{jheppub}
\usepackage{amsmath}
\usepackage{graphicx}
\usepackage[config=altsf]{subfig}
\usepackage{feynmf}
\usepackage{array}
\usepackage{slashed}
\usepackage{afterpage}
\usepackage{appendix}
\usepackage{multirow}
\usepackage{tikz}

\usepackage{bigfoot}
\DeclareNewFootnote[para]{B}[alph]
\DeclareNewFootnote{default}
\newcommand{\bear}{\begin{array}}
\newcommand{\ear}{\end{array}}

\newcommand{\beq}{\begin{eqnarray}}
\newcommand{\eeq}{\end{eqnarray}}
\newcommand{\beqa}{\begin{eqnarray}}
\newcommand{\eeqa}{\end{eqnarray}}

\def\OMIT#1{{}}
\newcommand{\lsim}{\mathrel{\rlap{\lower4pt\hbox{\hskip1pt$\sim$}}
    \raise1pt\hbox{$<$}}}         
\newcommand{\gsim}{\mathrel{\rlap{\lower4pt\hbox{\hskip1pt$\sim$}}
    \raise1pt\hbox{$>$}}}         
\def\simlt{\stackrel{<}{{}_\sim}}

\def\Br{{\rm Br}}

\usepackage[utf8]{inputenc}

\title{ Experimental Considerations Motivated by the Diphoton Excess at the LHC}

\author[a]{Prateek Agrawal,}
\author[b]{JiJi Fan,}
\author[a]{Ben Heidenreich,}
\author[a]{Matthew Reece,}
\author[a]{Matthew Strassler}

\affiliation[a]{Department of Physics, Harvard University, Cambridge, MA 02138, USA}
\affiliation[b]{Department of Physics, Brown University, Providence, RI 02912}

\abstract {
We consider the immediate or near-term experimental opportunities offered by some scenarios that could explain the new diphoton
excess at the LHC.
If the excess is due to a new particle $X_s$ at 750 GeV, additional new particles are required, providing further signals.
If connected with naturalness, the $X_s$ may be produced in top partner decays. Then a
$t'\bar t'$ signal, with $t'\to t X_s$ and $X_s\to gg$ dominantly,
might be discovered by reinterpreting 13 TeV SUSY searches in multijet
events with low MET and/or a lepton. If $X_s$ is a bound state of
quirks, the signal events may be accompanied by an unusual number of
soft tracks or soft jets.   Other resonances including dilepton and
photon+jet as well as dijet may lie at or above this mass, and
signatures of hidden glueballs might also be observable.  If the
``photons" in the excess are actually long-lived particles decaying to
photon pairs or to electron pairs, there are opportunities for
detecting overlapping photons and/or unusual patterns of apparent
photon-conversions in either $X_s$ or 125 GeV Higgs decays. There is
also the possibility of events with a hard ``photon'' recoiling
against a narrow isolated HCAL-only ``jet'', which, after the jet's energy
is corrected for its electromagnetic origin, would show a peak at 750
GeV.}
\preprint{\today}

\begin{document}
\maketitle

\section{Introduction}
\label{sec:intro}

Recently both ATLAS and CMS reported the first physics results from
Run 2.  One interesting small excess has been found in the diphoton
invariant mass spectrum around 750 GeV, with $\sim 10-20$ possible signal events
 and a global
significance of $\sim 2$  standard deviations at both CMS and
ATLAS~\cite{ATLAS-CONF-2015-081, CMSdiphoton}.  To determine that the
excess is from a real signal will require more integrated luminosity, but in the meantime
we may view the excess as a possible clue, one which might point us to other phenomena.  These phenomena can be sought, and discoveries potentially made, using existing or near-future data.

In this paper we will assume that the excess is due to a particle $X_s$ with mass  750 GeV, and we will discuss the distinctive experimental signatures of three simple new physics scenarios in which it might fit.  These include an obvious model with a pseudoscalar coupled to fermionic top partners, a scenario in which the ``photons'' to which $X_s$ decays are actually boosted and long-lived particles decaying to photons or $e^+e^-$, and the possibility that $X_s$ is a bound state of quirks (particles confined by a gauge group with no light matter.)  Each of these leads to non-obvious questions that can potentially be answered by careful examination of events with $X_s$ candidates and exploration of other data samples.

First we briefly summarize some simple facts about the current data.  It is perhaps surprising that there is an excess in Run 2 data that was not seen in Run  1 searches for diphoton resonances, which were conducted by CMS \cite{Khachatryan:2015qba,CMS:2015cwa} using 19.7 fb$^{-1}$ of data and by ATLAS \cite{Aad:2015mna} using 20.3 fb$^{-1}$.  For $gg\to X_s$ the sensitivity is only marginally higher, while for $q\bar q\to X_s$ it is reduced.
Denoting the parton luminosity ratio for a given initial state by
$R_i^{(13/8)}$, the relative sensitivity of Run 2 to Run 1 is
roughly\footnoteB{See \url{http://www.hep.ph.ic.ac.uk/~wstirlin/plots/plots.html}.}
\beq
\frac{(S/\sqrt{B})_2}{(S/\sqrt{B})_1} = \frac{R_i^{(13/8)}}{\sqrt{R_{q{\bar q}}^{(13/8)}}} \sqrt{\frac{L_2}{L_1}} \sim \begin{cases} 1.2 & {\rm if}~i = gg \\ 0.7 & {\rm if}~i = q{\bar q} \end{cases}
\eeq
for ATLAS and slightly less for CMS.  More sensitivity could be achieved with an improved analysis, but apparently the new analyses are not significantly different from those at 8 TeV. We therefore view $q\bar q\to X_s$ as unlikely.

Of course a much higher jump in the production rate could be achieved if $X_s$ is produced in the decay of a higher-energy particle, or is one of a pair of particles, so that the typical energy of the event is far above 750 GeV.  But in this case the events would generally have large additional energy, such as high $S_T$, high missing energy, hard jets, or some other feature that would make them look different from background.  Indeed, had this been the case, it would perhaps have been easier to confirm a signal on the basis of current data. ATLAS reports, however, that the events in the $X_s$ peak do not look different from the $\gamma\gamma$ background.  We therefore view this as unlikely.

We will focus therefore on $gg\to X_s$, loosely speaking in the case
of quirks.  To explain the excess at 13 TeV while remaining consistent
with the 8 TeV data, we will assume a diphoton rate in the range $(3 -
6)$
fb at 13 TeV~\cite{ATLAS-CONF-2015-081,
  CMSdiphoton}.\footnoteB{Note Added: In the
  previous version, the range 2--4 fb was taken; accounting for
efficiencies, ATLAS and CMS results from 8 and 13 TeV, and possible
isolation subtleties, we think the higher range is more
representative.  This change propagates into our numbers and Figure 1
but does not affect any of our conclusions.}
We will not assume the $X_s$ width is large, as statistics is too low to determine it, but we will consider this possibility.

Importantly, the $X_s$ cannot be the only new particle; more
generally, we cannot obtain the required gluon and photon couplings
using only the known particles of the Standard Model.  Since all such
particles have mass below 375 GeV, any loop graph of a Standard Model
particle $f$ that couples $X_s$ to photons necessarily has an
imaginary part, allowing a tree-level $X_sf\bar f$ decay that will
make $\Br(X_S\to \gamma\gamma)$ tiny.  Moreover the loop is suppressed
for $m_f\ll m_{X_s}$.  With over ten $X_s\to\gamma\gamma$ events, the
number of $Xf\bar f$ events is huge, and there is no way to hide such
a 750 GeV $f\bar f$ resonance in existing data
\cite{CMSdijet,Aad:2015fna}.  Therefore additional particles must be added, and in what follows we will consider how they might be found. 

The organization of this paper is as follows. In section~\ref{sec:simple}, we consider the simplest and most obvious model which could fit the observed excess, that of a new (pseudo)scalar coupling to gauge bosons via  effective operators of the form $X_s F^2$ or $X_s F\tilde{F}$. These couplings can be generated by integrating out heavy vector-like fields which could be top partners in a natural scenario. In section~\ref{sec:fake}, we consider several models where a different particle fakes a single photon in the detector, such as a light axion decaying to two unresolved photons or a dark photon decaying to $e^+ e^-$. In the former case, the observed resonance near 750 GeV is the scalar partner of the axions, decaying via $s\to a a$, whereas in the latter it is a (pseudo)scalar with an effective coupling $X_s F_D^2$ or $X_s F_D\tilde{F}_D$ to the dark photons. In section~\ref{sec:quirks}, we consider a model where the resonance is ``quirkonium'', a bound state of particles charged under a hidden confining gauge group with no light flavors. We conclude in section~\ref{sec:conclusions}.


\section{A Simple Natural Approach} \label{sec:simple}

\subsection{The simplest effective action}
\label{sec:everyone'smodel}

The obvious way to describe the observed excess is to make $X_s$ a scalar $s$ or pseudoscalar $\eta$ coupled to gauge boson by dimension-five operators, as in
\begin{equation}
  \label{modelA}
\kappa_G  \eta G_{\mu\nu}^a \tilde{G}^{a\mu\nu} + \kappa_\gamma  \eta F_{\mu\nu} \tilde{F}^{\mu\nu},
\end{equation}
where $G, F$ are $SU(3)_c$ and $U(1)_{\rm em}$ field strengths with
$a=1,\dots,8$ and $(\tilde{G}, \tilde{F})^{\mu\nu} = \frac12 \epsilon^{\mu\nu\alpha \beta} (G_{\alpha\beta}, F_{\alpha\beta})$. Here $\eta F_{\mu\nu} \tilde{F}^{\mu \nu}$ is a linear combination of the electroweak operators $\kappa_W \eta W^{\mu\nu; A} \tilde{W}_{\mu\nu}^A$ and $\kappa_B \eta B^{\mu \nu} \tilde{B}_{\mu\nu}$ where $W, B$ are the field strengths of $SU(2)_w$ and $U(1)_Y$ and $A =1,2,3$. 

For a scalar $s$, the discussion below follows almost unchanged, except for a slight change in cross-section.  A model-building issue is that one must explain why mixing of $s$ with the Higgs $h$ is small and why the scalar is natural.  To keep things simple we will focus on the pseudoscalar, but we emphasize that our phenomenological remarks are identical or only slightly modified in the scalar case.

The only definite prediction in such a model is the obvious one from
the electroweak theory: $\eta\to Z\gamma$ and $\eta\to ZZ$ cannot be
much smaller than $\eta\to\gamma\gamma$, and $\eta\to W^+W^-$ is a
possibility and an important diagnostic.  If the hypercharge term
dominates the decays, then the ratios of electroweak final states are
$\gamma\gamma:Z\gamma:ZZ:WW = c_W^4:4c_W^2s_W^2:s_W^4:0$ (where
$s_W,c_W$ are $\sin\theta_W$ and $\cos\theta_W$), while if $SU(2)$
dominates, the ratio is $s_W^4:4c_W^2s_W^2:c_W^4:2$.  Of course a linear combination is possible, but three ratios of branching fractions are determined by one parameter $\kappa_W/\kappa_B$, so there are two clear predictions, albeit ones that will not quickly be verified.

Such predictions could be violated if operators involving both $\eta$
and the SM Higgs $H$, with or without gauge bosons, can be generated.
This would imply 
renormalizable interactions that link $\eta$ and $H$, so it would be
very interesting were these predictions to fail.

Now let us consider how this model might arise.  Clearly Eq.~\ref{modelA} can be realized from a renormalizable perturbative theory by inducing these operators via loops of one or more heavy colored charged particles $f$ whose masses arise from something other than the Higgs, and are therefore little constrained by Higgs measurements.
After integrating out the heavy new fermions, we obtain 
\begin{eqnarray}
\kappa_G&=& \frac{\alpha_s }{4\sqrt{2}\pi } \sum_f  \frac{y_fN_f t_f}{m_f}, \\
\kappa_\gamma&=&\frac{\alpha }{4\sqrt{2}\pi } \sum_f  \frac{y_fN_c N_f Q_f^2}{m_f}, 
\end{eqnarray}
where the summation runs over all types of fermions, $y_f$ is the Yukawa coupling to $\eta$, $N_f$ is the number of flavors, $t_f$ is the Dynkin index of the colored fermions under $SU(3)_c$ (normalized to $1/2$ for fundamental representation) and $Q_f$ is the electric charge of the fermion. The branching ratio of $\eta$ to two photons is given by 
\begin{equation}
{\rm Br}(\eta \to \gamma \gamma) \approx \frac{\left(\alpha\sum_f \frac{ y_fN_c N_f Q_f^2}{m_f}\right)^2}{8\left(\alpha_s\sum_f\frac{y_f N_ft_f}{m_f}\right)^2}
\end{equation}
For $N_f$ copies of an $SU(2)$ singlet quark $t^\prime$ of charge 2/3, plus its vector-like partner $\bar{t}^\prime$, $ {\rm Br} (\eta \to \gamma \gamma) \approx 0.4 \%$. For $N_f$ copies of $SU(2)$ doublets $Q^\prime = (t^\prime, b^\prime)$ of charge $(2/3,-1/3)$ and their vector-like partners, $ {\rm Br} (\eta \to \gamma \gamma) \approx 0.15 \%$. These two choices will be our benchmark models in the following discussions. We will also assume that heavy quarks are all degenerate in mass for simplicity. 

The production cross section of $\eta$, when we add the vector-like color triplets with a common $y_f$ and $m_f$, is approximately
\begin{equation}
\sigma_{pp\to \eta} \approx 400\,{\rm fb}\, \left(\frac{y_f N_f^{\rm eff} v}{ m_f}\right)^2
\end{equation}
where $v=246$ GeV is the Higgs vev.\footnoteB{The prefactor of 400 fb is the cross-section for a 750 GeV Standard Model Higgs in the limit that $m_t\to \infty$.   We computed this in
  Madgraph with the Higgs effective theory (heft) model file~\cite{Alwall:2014hca} and assigned a $K$ factor $\sim 2$ (for 14 TeV, the K factor for $ggh$ of a 750 GeV Higgs is around 1.8~\cite{Djouadi:2005gi}).}
In models with singlet quarks $t'_i$, $N_f^{\rm eff} = N_f$, while in models with doublets $Q^\prime_i$, $N_f^{\rm eff} = 2N_f$.

In the case of $t^\prime$ or $Q^\prime$, this can be translated into a bound on the combination of the parameters,
\begin{equation}
m_f = \begin{cases}  y_f N_f^{\rm eff} \, (245 \mathrm{\ GeV})  / (1.4 \to 1.9)  &   t^\prime \\
y_f N_f^{\rm eff} \,(245 \mathrm{\ GeV}) / (2.2 \to 3.2) & ( t^\prime, b^\prime) \end{cases}\,,
\label{eq:masstrivialmodel}
\end{equation}
where the lower (higher) number in the denominator corresponds to $3$
($6$) fb for $\sigma(pp\to X_s$).
In cases with $t^\prime$ decaying to  $t + \eta$ or to $t (b) +$ SM
gauge bosons, the new quarks have to be heavier than $\sim 1$ TeV, as
we will discuss below, and so $y_f N_f^{\rm eff} \gtrsim 6$.
To
keep the theory perturbative, (i.e.\ the one-loop correction to $y_f$
smaller than $y_f$),\footnoteB{Note added: In a previous version we
used a slightly stronger criterion; Figure 1 now has a slightly weaker
perturbativity constraint, but of course this is not a sharp boundary
in any case.} we need $y_f^2 N_c N_f \lesssim 16 \pi^2$. 

In general one can obtain any values of $\kappa_G, \kappa_\gamma$ desired, by including multiple particles with different masses, hypercharges, etc., and including colorless particles. Therefore one must view the $\kappa_G, \kappa_\gamma$ as free parameters, only requiring that they not be so large as to violate perturbativity in the effective theory.

Since $\kappa_G$ is induced at one loop, this type of model tends to
give $\eta$ a width below 1 GeV, and does not immediately explain how
a resonance could have a width of tens of GeV, as ATLAS's excess seems
tentatively to suggest.  However, the model is easily modified. For
instance, a renormalizable interaction of $\eta$ with invisible
particles could greatly increase the width.  Meanwhile a moderate
increase in both $\kappa_G$ and $\kappa_\gamma$ (for instance by
doubling the number of heavy quarks and adding several heavy leptons)
would allow the diphoton signal to stay unchanged,  without creating a
750 GeV dijet signal so large that it would be excluded by CMS at Run
1 \cite{CMSdijet}.

\subsection{A natural scenario and its predictions}

Could a simple model of this type be related in some way to
naturalness, and if so, what might one search for?  One possibility is
that we could identify $t^\prime$ and/or $(t^\prime, b^\prime)$ as the
top partners, and $\eta$ as a pseudoscalar arising in {\it e.g.}\ a
little Higgs model~\cite{ArkaniHamed:2001nc}. In this case we expect them to be light enough to
be produced at the LHC.  

In many models the $t^\prime$ ($b^\prime$) decay dominantly to $Wb,
tZ, th$ ($Wt, bZ, bh$), and most searches for $t'$ ($b^\prime$) have
assumed this. 
The Run 1 constraint on $t^\prime$ of $\sim 800$
GeV~\cite{Khachatryan:2015oba} has just been pushed up to close to a
TeV~\cite{ATLASCMS} with Run 2 data.  We will typically require multiple copies
of $t^\prime$ (and $b^\prime$ in the second benchmark) to obtain an
acceptable rate, increasing the signal and therefore the inferred mass
limit.  

But the presence of the $\eta$ offers the possibility that $t'\to
t\eta$ (and $b'\to b\eta$) will be the dominant decay
channel~\cite{Kilian:2004pp,Kilian:2006eh}.  For
simplicity let us consider singlet $t'$ quarks first.
Suppose the model has a $Z_2$ ``T-parity'' \cite{Cheng:2004yc} under
which $t'\to -t'$. Exact parity would make $t'$ stable, so the parity
must be violated, for instance by a small coupling $\eta
\bar{t}^\prime t_R$ where $t_R$ is the SM right-handed top. This will
induce $t'\to t\eta$ decays. Kinetic mixing between $t^\prime$ and
$t_R$ appears at one-loop order.  This allows $t^\prime\to tZ, th$,
but at a rate suppressed compared to $t^\prime \to t \eta$, as long as
$m_{t'}-m_t>m_\eta=750 \mathrm{\ GeV}$.  (For lower $m_{t'}$ the
on-shell $tZ,th$ decays are favored over $t'\to tjj$ via an off-shell
$\eta$.)  Similarly, in model with doublets $t',b'$ and an approximate
T-parity, either or both $t^\prime,b^\prime$ will dominantly decay to
$t\eta, b\eta$.

Pair production of $t^\prime$ ($b^\prime$) and the subsequent decays $t(b)^\prime \to t(b) \eta, \eta \to \gamma\gamma$ will contribute to the diphoton bumps at ATLAS and CMS, with the number of events being 
\begin{eqnarray}
N_{f}^{\rm eff} \sigma_{pp \to t^\prime t^\prime} \times 2   {\rm Br} (\eta \to \gamma\gamma) {\cal L}\,.
\end{eqnarray}
Here $\sigma_{pp \to t^\prime t^\prime}$ is the cross section to
produce a single $t^\prime$. We take the branching fraction of new
quarks decaying to $\eta$ plus a SM quark to be nearly one.  These
high-energy diphoton + multijet events would be spectacular and, if
present in the region of the diphoton excess, could not fail to be
noticed.  Therefore we assume that none have been seen and we
conservatively take the expected number of such events to be 2 or
fewer.\footnoteB{Note added: this may be too strong a constraint given the acceptance of the ATLAS diphoton search.  Weakening the constraint allows for a potentially larger $t\bar tjjjj$ signal described below.}

\begin{figure}[ht]
  \centering
  \includegraphics[width=0.45\textwidth]{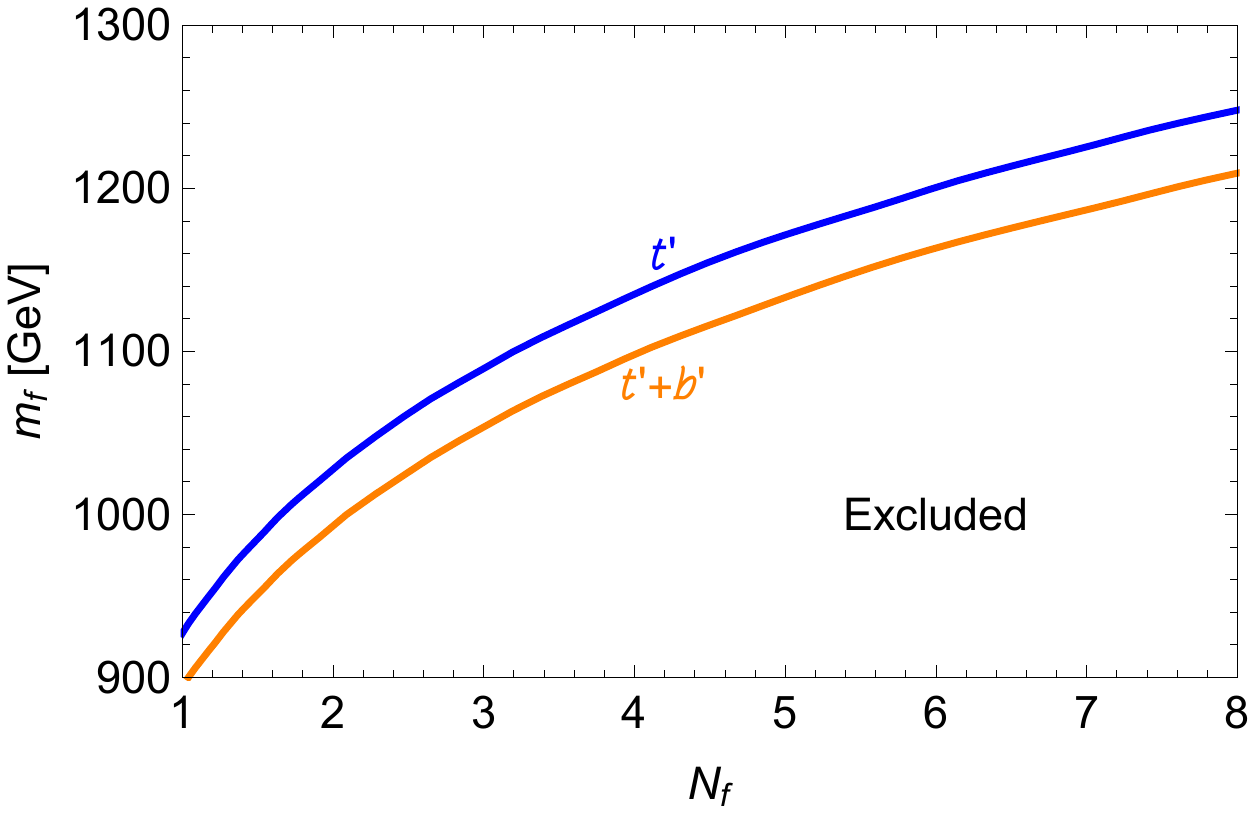} \quad   \includegraphics[width=0.43\textwidth]{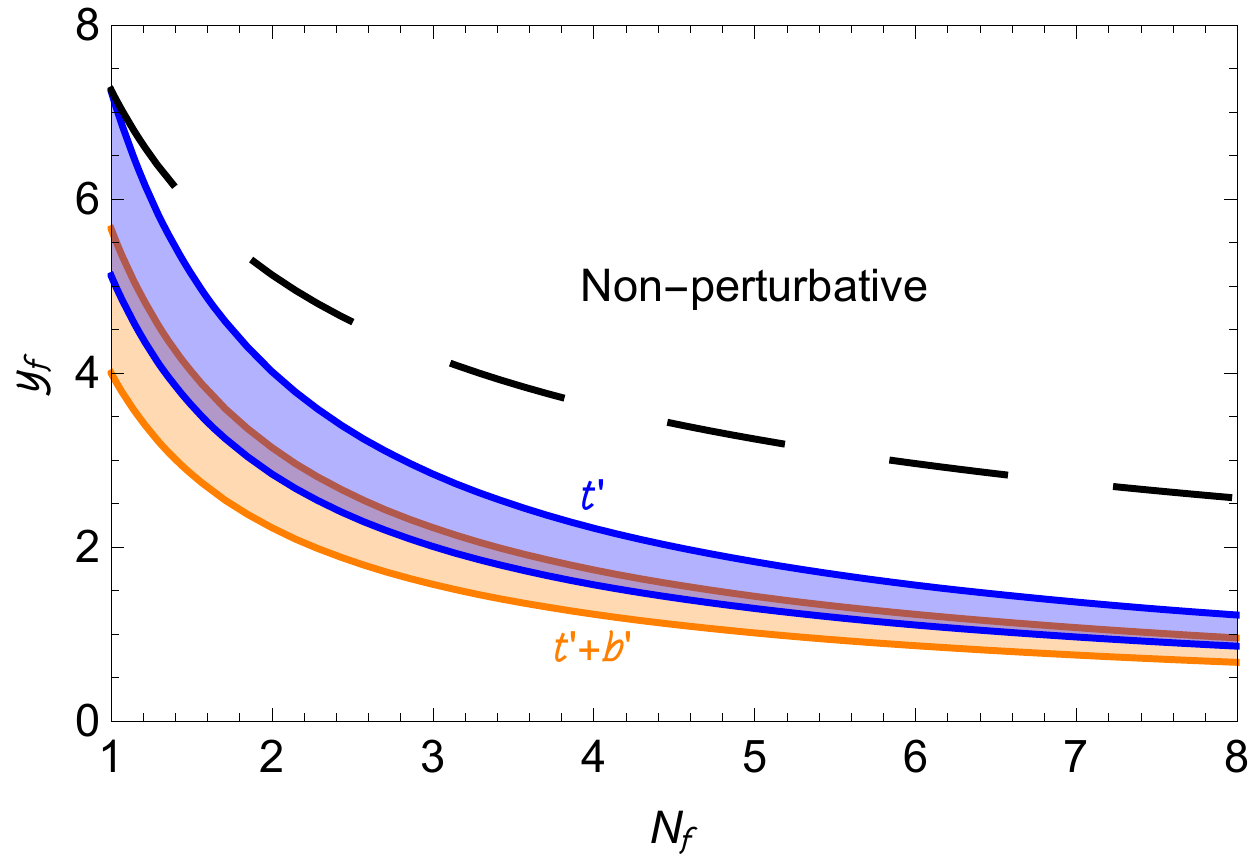}
  \caption{Left: Lower bound on the single $t^\prime$ (blue) and doublet $t^\prime,b^\prime$ mass $m_f$ (orange) as     a function of $N_f$, from requiring the
    expected contribution of $t^\prime \to t \eta \to t\gamma\gamma$ to the
    diphoton bump is less than two events at ATLAS Run 2 (with luminosity
    of 3.2 fb$^{-1}$.) Right: Taking the minimal $m_f$ for each $N_f$ from the left plot, the bands show the required value of $y_f$ needed to match the diphoton signal, from Eq.~(\ref{eq:masstrivialmodel}). The blue (orange) band is for $t^\prime$  $(t^\prime, b^\prime)$ quarks.   The rough perturbativity bound on $y^2N_f$ is shown as a dashed line.
      }
  \label{fig:everyonemodel}
\end{figure}

This requirement bounds the masses of $t^\prime$ and $b^\prime$ as a
function of $N_f$.  This is shown in the left panel of
Fig.~\ref{fig:everyonemodel}. Then, taking the smallest allowed $m_f$
for each $N_f$, we can determine the $y_fN_f$ in
Eq.~(\ref{eq:masstrivialmodel}) needed to get the required
cross-section for $gg\to\eta\to\gamma\gamma$. This is shown in the
right panel of Fig \ref{fig:everyonemodel}.   We see that we are approaching
the edge of perturbativity, though still plausibly below it if $N_f>1$, which in turn implies
$t',b'$ must have masses near or above 1 TeV.
beyond $t',b'$ must have masses above 1 TeV.
Naturalness considerations would then
encourage us to keep the masses close to their lower bound.

Our main point is that this scenario leads to a signal that could be found in current LHC data.  If $t'\to t\eta$ is dominant, then $t'$ pair production leads to $t\bar t\eta\eta$, with the majority going to $t\bar t$ plus four very hard jets.  
Given the constraint that there are at most two $t\bar t jj\gamma\gamma$ events at the $\eta$ resonance, there could be as many as $\sim 100$ $t\bar tjjjj$ events with a leptonic top decay in the data.  Searches sensitive to this signal include those for many jets and small missing energy (as in Run 1~\cite{Aad:2013wta} and recently in Run 2~\cite{ATLAS-CONF-2015-077}) and would also include searches for single lepton + jets and low missing energy (as suggested in~\cite{Lisanti:2011tm}.)  We therefore propose that searches of this type be reinterpreted, and perhaps further optimized, for signals from $t'\to t\eta$ pairs.

More quantitatively, the rate in question is a good fraction of that expected for pair-produced 1.2 TeV gluinos decaying to the same final state. Since a gluino at this mass certainly could be seen in the data already, existing data may already be sensitive to this $t'\to t\eta$ signal. A recast \cite{Anandakrishnan:2015yfa} of the ATLAS multijet search at Run 1~\cite{Aad:2015lea} shows that such a signal for $N_f=1$ is only excluded up to 800 GeV and $\eta$ mass up to 500 GeV. The LHC experiments should be able to use existing Run 2 data to probe this decay topology for $N_f$ $t^\prime$ quarks with mass above a TeV.

\section{Models with fake photons} \label{sec:fake}

As numerous authors have emphasized
\cite{Dobrescu:2000jt,Chang:2006bw,Toro:2012sv,Draper:2012xt,Ellis:2012sd,Ellis:2012zp,Curtin:2013fra}, we should be cautious
about whether experimental ``photons'' are truly photons. False
photons can arise from a highly boosted object that decays (possibly
at a displaced location) to multiple photons, which then hit the
electromagnetic calorimeter (ECAL) at essentially the same
place.\footnoteB{We thank
Lian-Tao Wang for reminding us of this possibility.}
They can also arise from a boosted and displaced object decaying to
$e^+e^-$ pairs from a displaced vertex, causing their tracks to be
lost and/or appear to be from a photon conversion.
In this section we will focus on the first case, considering a light pseudoscalar $a$ that decays to two collimated photons, though we will also
briefly discuss the second.

\subsection{Collimated photon pairs}

Small opening angles and long lifetimes could both enhance the rate of $a \to \gamma\gamma$ decays reconstructed as single photons. ATLAS and CMS have somewhat different capabilities for detecting
photons and distinguishing them from other possible electromagnetic
objects like hypothetical light pseudoscalars. The CMS ECAL is situated $R = 1.3$ meters in radius from the beam and has segmentation
$\Delta \eta \times \Delta \phi = 0.0174 \times
0.0174$ \cite{Khachatryan:2015iwa}. The ATLAS ECAL lies at $R = 1.5$ meters and has
$\Delta \eta \times \Delta \phi = 0.025 \times 0.025$
\cite{Aad:2009wy,Aad:2010sp}. The first layer of the ATLAS ECAL
consists of 3 to 5 radiation lengths in the form of narrow strips of
fine granularity as small as $\Delta \eta = 0.003$ (and ranging up to
$0.006$, depending on $\eta$), which plays a role in $\pi^0$ rejection
\cite{Aad:2010sp}. In the barrel region ATLAS photon reconstruction
uses a $3 \times 5$ or $3 \times 7$ cluster of neighboring cells,
while in the endcaps it uses a $5 \times 5$ cluster. A full detector
simulation would probably be required to get a realistic understanding
of when the ATLAS and CMS photon ID procedures call an $a \to \gamma
\gamma$ a photon. But we can still learn something interesting and suggestive
by looking at how the typical separation between the two photons depends on the mass, lifetime
and boost of the $a$.

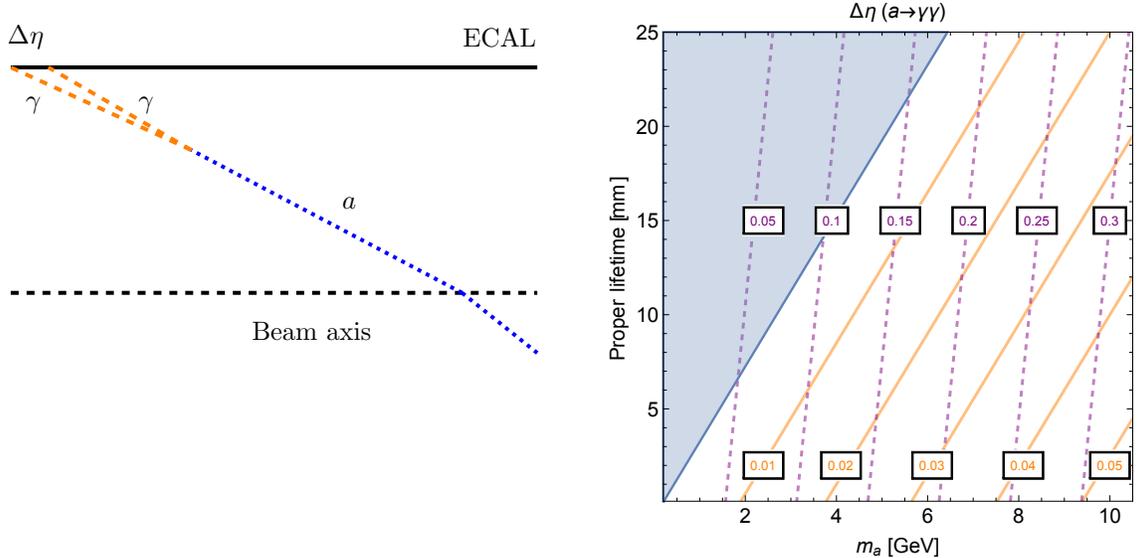
\begin{figure}[h]
\begin{center}
\begin{tikzpicture}[line width=1.5 pt]
\draw[dashed] (-4,0)--(3.0,0);
\draw (-4,3)--(3.0,3);
\draw[white] (-4,-3.5)--(3.0,-3.5);
\draw[dotted,blue] (2,0)--(-1.6,1.9);
\draw[dotted,blue] (2,0)--(3.0,-0.8);
\node at (0.5,1.2) {$a$};
\node at (0.0,-0.5) {Beam axis};
\node at (2.5,3.4) {ECAL};
\draw[dashed,orange] (-1.6,1.9)--(-4.0,3);
\node at (-3.7,2.5) {$\gamma$};
\draw[dashed,orange] (-1.6,1.9)--(-3.5,3);
\node at (-2.2,2.5) {$\gamma$};
\node at (-3.8,3.4) {$\Delta \eta$};
\end{tikzpicture}
\qquad \includegraphics[width=0.45\textwidth]{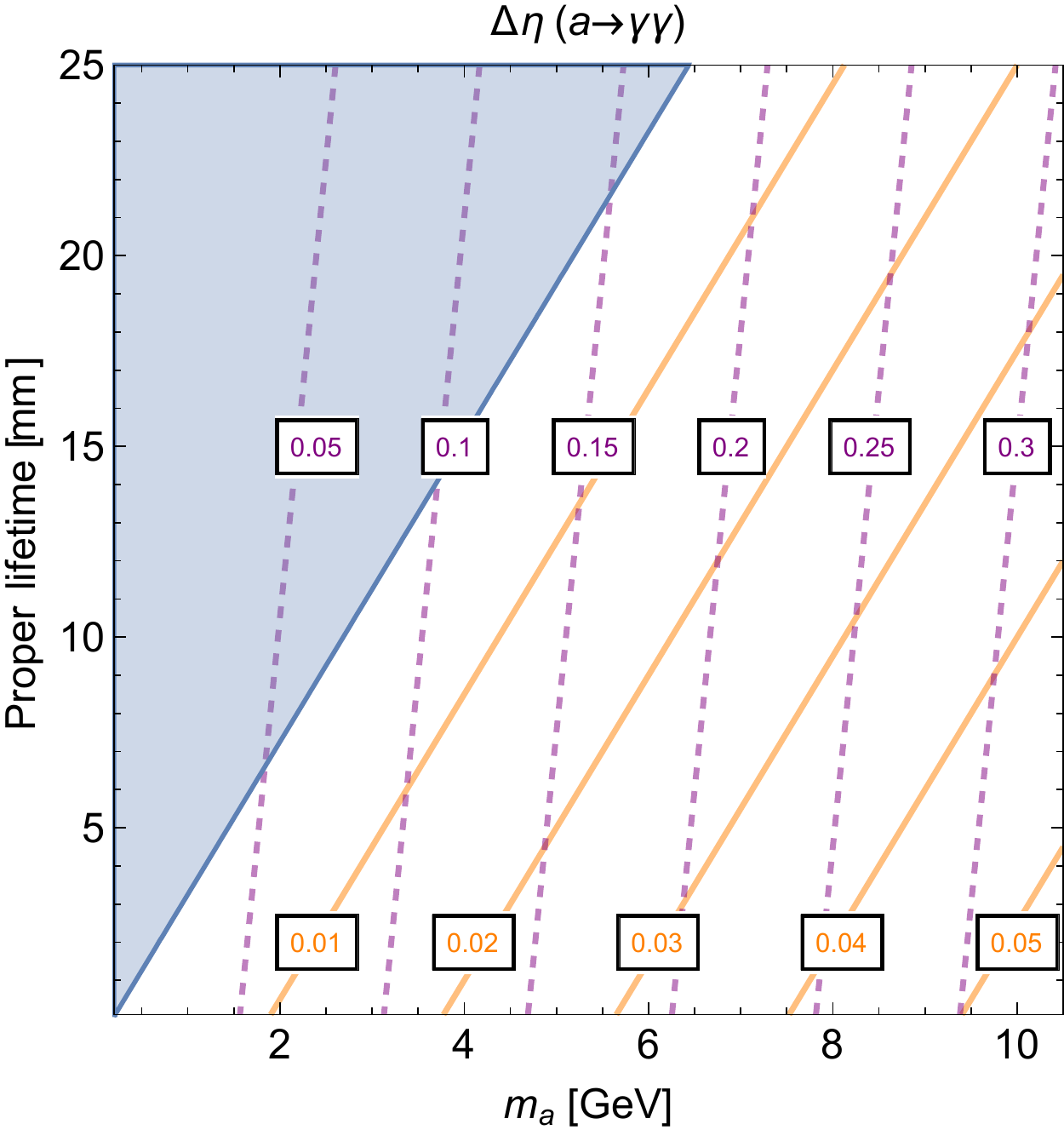}
\end{center}
\caption{{\bf Left:} an illustration of how a long lab-frame lifetime can reduce the angular separation $\Delta \eta(\gamma,\gamma)$ as measured in the ECAL. {\bf Right:} The angular separation $\Delta \eta$ between two photons produced in the decay $a \to \gamma\gamma$ where the $a$ has a large boost. The orange solid contours correspond to $E_a = 375~{\rm GeV}$, as in the decay of a 750 GeV resonance, whereas the purple dashed contours correspond to $E_a = m_h/2 = 62.5$ GeV. The vertical axis shows the proper lifetime $\widehat{c\tau}$ in meters; we assume that the particle decays at a finite radius $L = \gamma \widehat{c\tau}$. In the blue shaded region, the particle produced in the decay of the 750 GeV resonance reaches the ECAL (assumed to be at 1.5 meter radius, appropriate for ATLAS) before it decays. }
\label{fig:deltaeta2}
\end{figure}

We will adopt the notation $c\tau$ for the proper lifetime of a particle,
i.e. $\tau = \Gamma^{-1}$, the parameter that determines the
distribution of actual proper decay times. 
For the
proper time at which a {\em particular} particle decays, we will
reserve the notation $\widehat{c\tau}$. The opening angle of the $a$
decay is $\theta \approx 2m_a / E_a$. The lab-frame lifetime is
$\widehat{c \tau} \gamma \approx 2\widehat{c \tau}/\theta$. So a
larger boost has two implications: one is a smaller opening angle. The
second is that the particle is likely to travel longer before
decaying, which places it closer to the ECAL and leads to a smaller
separation between the two photons when they arrive at the detector.
Consider the simple case of a particle $a$ propagating vertically
upward at $\eta = 0$ and decaying to two photons after traversing a
distance $\gamma \widehat{c\tau}$. Assume that these photons spread
dominantly in the $z$-direction.
Then they will reach the calorimeter at $z \approx \frac{1}{2} \theta
(R - \gamma \widehat{c \tau})$, so we have
\beq
\Delta \eta \approx \theta \left(1 - \frac{\gamma \widehat{c \tau}}{R}\right) \approx \frac{2}{\gamma} \left(1 - \frac{\gamma \widehat{c \tau}}{R}\right).
\eeq
This angular scale is shown in Fig.~\ref{fig:deltaeta2}. We see that particle masses near or somewhat above 1 GeV and proper lifetimes of around a millimeter can give photons with small separation on the scale of the ECAL segmentation in a process $s \to aa, a\to \gamma\gamma$ with $m_s \approx 750~{\rm GeV}$. In the same region of parameter space, the angular separation of the photons in Higgs decays $h \to aa, a \to \gamma\gamma$ is significantly larger than an ECAL cell and unlikely to pass photon ID cuts. We will return to this point below in the context of a specific benchmark model, which allows us to look at the full distribution of angular separations.

\subsection{Resonance decaying to axions} 
\label{sec:BLMmodel}

Now we consider if there is a viable model that could potentially explain the excess. We consider first a simple model of \cite{Dobrescu:2000jt} in which $X_s$ production involves its mixing with the Higgs.  This model is viable, though with constraints that probably limit cross-sections to several fb. We also briefly consider models in which $X_s$ production occurs via loops of colored particles, which have no such constraints.
Along the way we identify specific experimental signals which are characteristic of these
classes of models.

Following \cite{Dobrescu:2000jt}, we add a complex singlet scalar $S$ and one heavy lepton $L$ to the SM. 
The interactions are
\begin{align}
  \mathcal{L}_{int}
  &=
  -\frac{\lambda_H}{2} (H^\dagger H)^2
  -\frac{\lambda_S}{2} (S^\dagger S)^2
  - \lambda_0 H^\dagger H S^\dagger S
  + M_H^2 H^\dagger H
  - M_S^2 S^\dagger S
  \nonumber \\&\qquad
  + ( y S L^c L
  + \epsilon M_S^2 S^2 + {\rm h.c.}) \ .
  \label{eq:full-lag}
\end{align}
In general we might have multiple vector-like fields coupling to $S$. The parameter $\epsilon$ breaks the global $U(1)$ symmetry associated with
$S$. (In \cite{Dobrescu:2000jt} the symmetry breaking was chosen to be a tadpole term; the exact choice has little impact on the phenomenology.) We assume that $S$ gets a VEV and mixes with the Higgs. We parametrize $S$ as
\begin{align}
  S=\frac{1}{\sqrt{2}} \left(s+f\right)e^{ia/f}
\end{align}
so that $s$ and $a$ both are  canonically normalized.
We integrate out $L$ and write the phenomenologically
relevant terms in the Lagrangian (keeping the leading behavior),
\begin{align}
\mathcal{L}_{int}
&=
- \lambda_0 v f h s
+\frac{s}{f}\partial_\mu a\, \partial^\mu a
+ \frac{\alpha q^2}{4\pi f}
a F_{\mu\nu} \tilde{F}^{\mu\nu}.
\end{align}
Here $q^2$ would be replaced by a more complicated factor if there were multiple particles in the loop or the mass of $L$ does not come entirely from $S$.

The $a$ boson decay width is
\beq
\Gamma(a \to \gamma\gamma) = \frac{m_a^3}{4\pi} \left(\frac{\alpha q^2}{4\pi f}\right)^2 = \frac{1}{0.74~{\rm cm}} q^4 \left(\frac{m_a}{1~{\rm GeV}}\right)^3 \left(\frac{1~{\rm TeV}}{f}\right)^2
\eeq 
Accounting for the large boost of the $a$ in $h\to aa$ and $s\to aa$ decays, we find
\begin{align}
h{\rm \ decays}:~\gamma c\tau &\approx 0.5~{\rm m} \left(\frac{f}{q^2~{\rm TeV}}\right)^2 \left(\frac{1~{\rm GeV}}{m_a}\right)^4. \\
S{\rm \ decays}:~\gamma c\tau &\approx 2.8~{\rm m} \left(\frac{f}{q^2~{\rm TeV}}\right)^2 \left(\frac{1~{\rm GeV}}{m_a}\right)^4.
\end{align}
The ECAL radius is about a meter.  To assure most $a$ bosons decay before passing through the ECAL, we simply need to take $f/q^2<600$ GeV or $m_a> 1.3$ GeV.

The Standard Model Higgs boson would have a 13 TeV production rate of about 750 fb if its mass were 750 GeV \cite{Dittmaier:2011ti}. Therefore
\begin{align}
  \sigma (p p \to s)
  &= 
   \left(\frac{\lambda_0 v f}{m_s^2 - m_h^2}\right)^2 \sigma(p p \to h(m_s))
 \approx 
  7.5\ \mathrm{fb} \left(\frac{\theta}{0.1}\right)^2,
\end{align}
where we have defined the small mixing angle
\beq
\theta \equiv \frac{\lambda_0 v f}{m_s^2 - m_h^2}.
\eeq
Both $s$ and the Higgs $h$ can decay to $aa$:
\begin{align}
\Gamma(h \to aa) &\approx \frac{\lambda_0^2 v^2 m_h^3}{32 \pi (m_s^2 - m_h^2)^2} \approx 0.04~{\rm MeV}\, \left(\frac{\lambda_0}{0.1}\right)^2, \\
\Gamma(s \to aa) &\approx \frac{m_s^3}{32 \pi f^2} \approx 4.0~{\rm GeV} \left(\frac{1~{\rm TeV}}{f}\right)^2.
\end{align}
The $s \to aa$ decay competes with the $s$ mixing through a Higgs and decaying to Standard Model final states,
\begin{align}
\Gamma(s \to {\rm SM}) &= \theta^2 \Gamma(h[750] \to {\rm SM}) \approx 2.5~{\rm GeV} \left(\frac{\theta}{0.1}\right)^2.
\end{align}
where $\Gamma(h[750] \to {\rm SM}) \approx 250~{\rm GeV}$ is the total width of a 750 GeV Standard Model Higgs boson \cite{Dittmaier:2011ti}.

\begin{figure}[h]
\begin{center}
\includegraphics[width=0.5\textwidth]{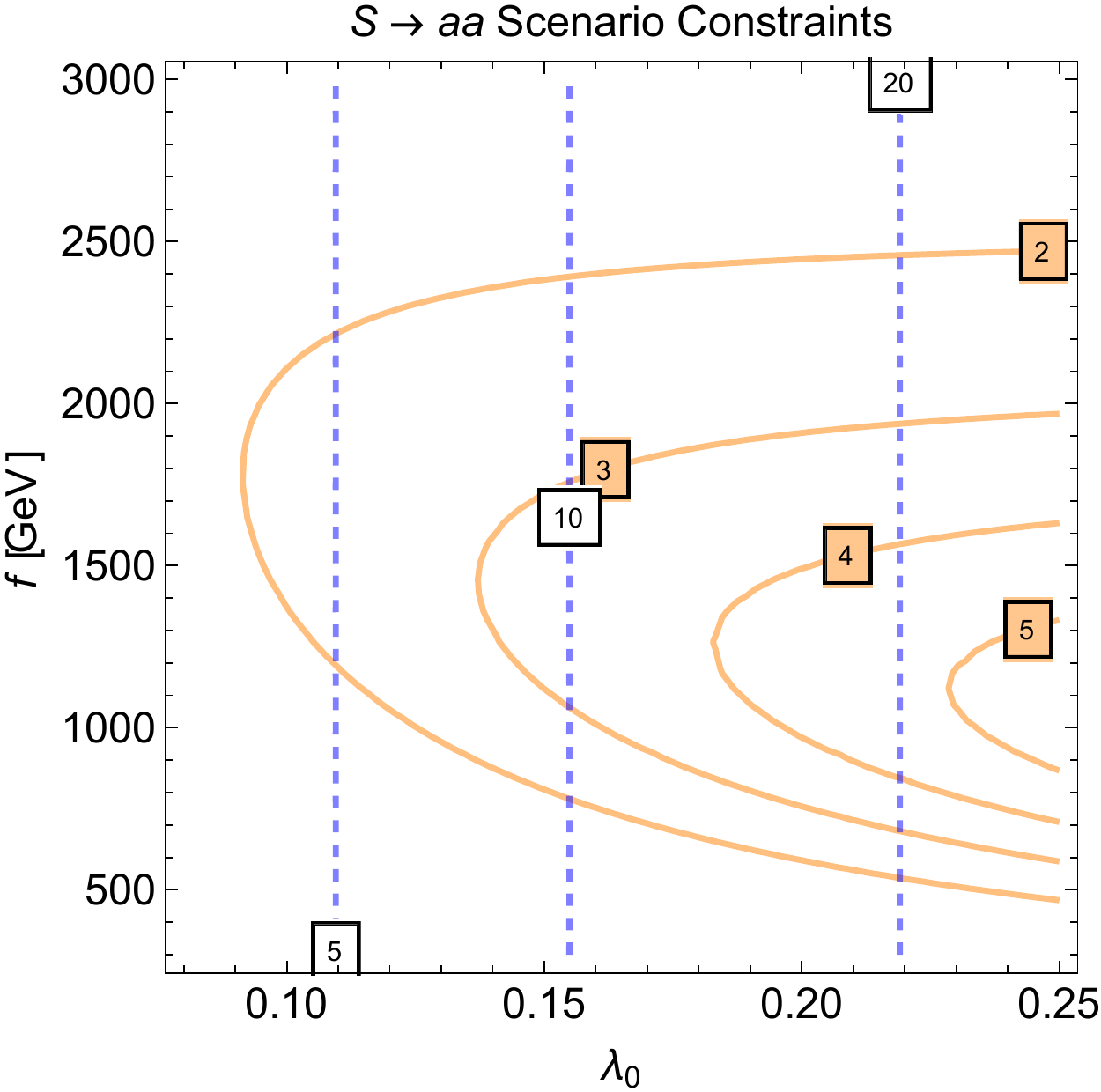}
\end{center}
\caption{Constraints on the scenario where a 750 GeV $s$ decays to
  $aa$ with $a \to \gamma\gamma$. The horizontal axis is the
  coefficient of $S^\dagger S H^\dagger H$ in the Lagrangian and the
  vertical axis is $f = \left<S\right>/\sqrt{2}$. The orange curves
  are contours of $\sigma(pp \to s) \times {\rm Br}(s\to aa)$ in fb.
  The blue dashed lines are contours of the exotic Higgs decay rate
  relative to the Standard Model Higgs to diphoton rate, $\Gamma(h \to
  aa)/\Gamma_{\rm SM}(h \to \gamma\gamma)$.  We see that if $s$
production requires Higgs mixing, fitting  the diphoton excess
necessarily entails a significant Br$(h\to aa)$.}
\label{fig:Saamodel}
\end{figure}

As shown in Figure \ref{fig:Saamodel}, there is no way to obtain a
suitable production $\sigma\times\Br$ for $pp\to s\to aa$ without a
substantial $\Br(h\to aa)$.
As a sample benchmark, we take $\lambda_0 = 0.23$, $f = 1200~{\rm GeV}$,
$q=1$ and $m_a = 2~{\rm GeV}$.\footnoteB{Note added: In the previous
  version we selected a benchmark with $\sigma(pp \to s) \times
  {\rm  Br}(s \to aa) \approx 2.5~{\rm fb}$. Here we take a benchmark that
lies within our updated range, with the consequence of a somewhat
larger Br($h\to aa$). }
For this choice we have, 
\begin{equation}
\begin{aligned}[c]
\theta &\approx 0.23 \\
\sigma(pp \to s) \times {\rm Br}(s \to aa) &\approx 5~{\rm fb} \\
\Gamma(h \to aa) \approx 0.2~{\rm MeV} &\approx 22.1\, \Gamma_{\rm SM}(h \to \gamma\gamma) \\
{\rm Br}(h \to aa) &\approx 5.1\% \\
\Gamma(s \to aa) &\approx 2.9~{\rm GeV}
\end{aligned} \qquad \qquad
\begin{aligned}[c]
\Gamma(s \to {\rm SM}) &\approx 3.8~{\rm GeV}\\
{\rm Br}(s \to aa) &\approx 0.43\\
c\tau(a) &\approx 1.3~{\rm mm}\\
c \tau \gamma(a)|_{\rm Higgs} &\approx 4.2~{\rm cm}\\
c \tau \gamma(a)|_s &\approx 25~{\rm cm}.
\end{aligned}
\end{equation}
Were most $h \to aa$ decays accounted for as $h \to \gamma\gamma$ events, this scenario would be strongly excluded by measured Higgs properties. We then have to demand that the $a$ bosons produced in Higgs decays are not too collimated, so that they don't pass photon ID cuts.

Conversely a search for 3 photon events at ATLAS~\cite{Aad:2015bua}
has excluded $h\to aa$ for mass $m_a> 10$ GeV and $\Br(h\to aa)\sim
3\times 10^{-4}$.  The sensitivity of the search was not shown below
10 GeV but was not yet dying off, so we presume a severe constraint
persists several GeV lower.  However, we think our benchmark point of
2 GeV is unlikely to be excluded, nor would we expect one at 4 GeV to
be excluded yet either.

A search for collimated photon pairs from $h \to aa \to 4 \gamma$
exists at ATLAS in the 7 TeV dataset \cite{ATLAS:2012soa}. This placed
bounds on the cases $m_a = 100, 200,$ and $400~{\rm MeV}$ at the level
of $\sigma \times {\rm Br} \simlt 100~{\rm fb}$. Given a total gluon
fusion Higgs production cross section of about 15 pb at 7 TeV, this
allows only a ${\rm Br}(h \to aa) \simlt 0.7\%$. However, the results
may be very sensitive to the $a$ mass. Unfortunately, further
experimental searches have not been carried out so far, though the
possible reach has been studied
\cite{Chang:2006bw,Toro:2012sv,Draper:2012xt,Ellis:2012sd,Ellis:2012zp,Curtin:2013fra}.  

\begin{figure}[t]
\begin{center}
\includegraphics[width=0.45\textwidth]{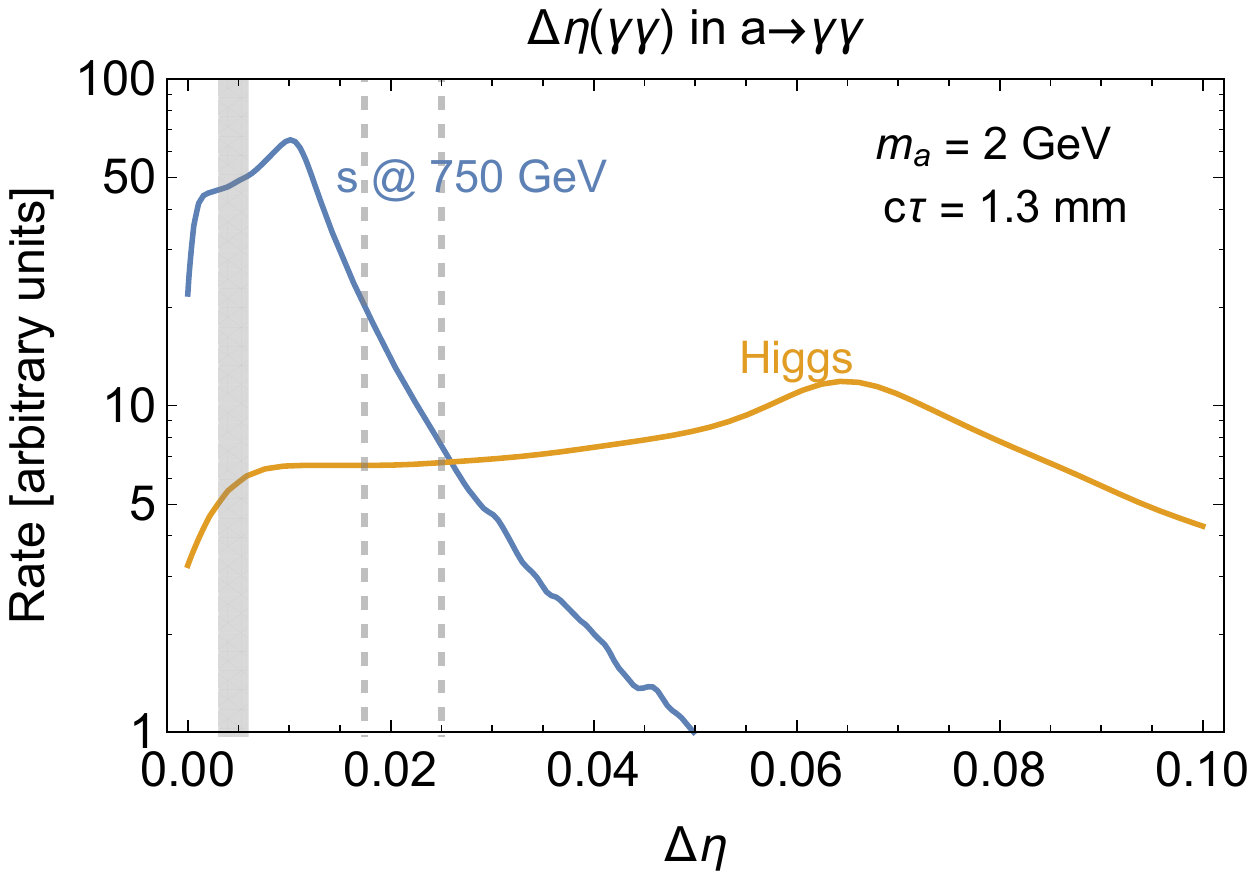}
\qquad
\includegraphics[width=0.45\textwidth]{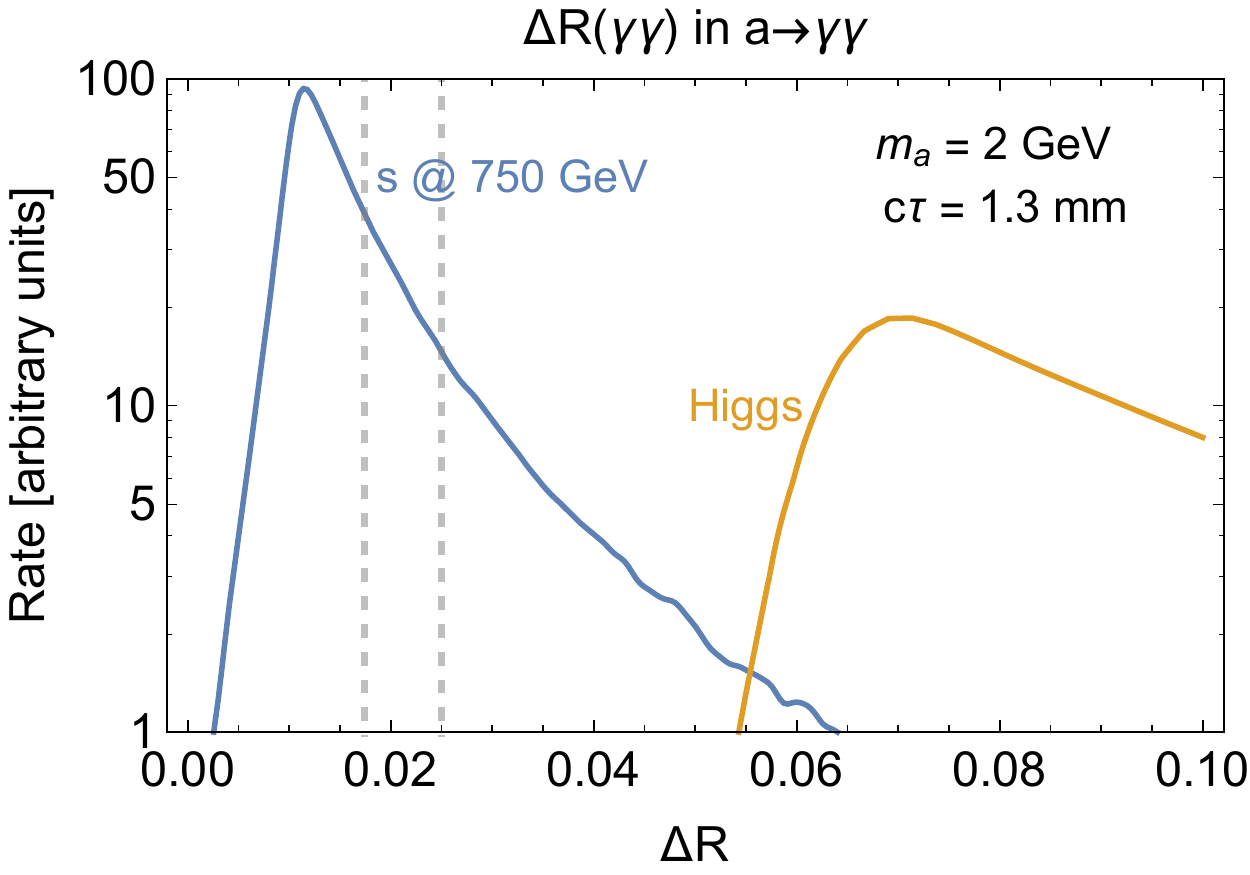}
\end{center}
\caption{For $m_a=2$ GeV and $c\tau=1.3$ mm, the angular separation
  $\Delta \eta$ and $\Delta R$ between two photons produced in the
  decay $a \to \gamma\gamma$ where the $a$ has a large boost. The blue
  curves correspond to the $s \to aa$ process with $m_s = 750~{\rm
  GeV}$, while the orange case is the less boosted decay $h \to aa$ of
the Higgs boson. The dashed vertical lines correspond to the angular
granularity 0.0174 and 0.025 of the CMS and ATLAS ECALs, respectively.
The shaded gray region corresponds to the fine segmentation in $\Delta
\eta$ available in the first layer of the ATLAS ECAL (which ranges
between 0.003 and 0.006 depending on $\eta$). We see that Higgs decays
have sufficient angular separation that the two photons together are
unlikely to be classified as a single well-identified photon, while in
the $s$ decay scenario the photons may be so collimated that they are
observed as a single photon.}
\label{fig:deltaeta}
\end{figure}

In Figure \ref{fig:deltaeta} we see that, for our benchmark point, the
distribution of opening angles between photons arising from $h\to aa$
and $s\to aa$ are significantly different.  Two effects go in the same
direction: the boost of the $s\to aa$ photons makes their opening
angle smaller than for $h\to aa$ even were $a$ to decay promptly, and
the longer lifetime in the lab frame due to the same boost reduces the
opening angle between the points where the photons impinge upon the
ECAL.  There are other more subtle effects as well.  Isolation cuts
may or may not cause a converted photon near a second (unconverted or
converted) photon to be rejected.  But for a highly boosted and highly
displaced $a$, it is both less likely for either photon to convert and
more likely that they overlap tightly in the ECAL.

The unusual patterns of overlapping photons and photon conversions
makes it impossible for us to model or predict how individual CMS and
ATLAS searches for photons would have responded to this signal.  It
may well be that the efficiencies of the various searches have been
quite different, depending on photon definitions and choices of photon
categories.\footnoteB{Perhaps this is a partial explanation of why observed $h\to \gamma\gamma$ signal strengths varied as much as they did.}  We can therefore do no more at this stage than encourage the ATLAS and CMS groups to take this possibility seriously and explore it thoroughly.

The width of the $s$ in our benchmark model is much larger than in the models of the previous section, but still quite a bit smaller than the tens of GeV currently suggested by ATLAS data.  However, because the ``photons'' in our model are actually photon pairs, and conversions are more likely as are overlapping showers, we expect that the resolution for the $s\to aa$ resonance is considerably worse than the 8 GeV that would be expected for a resonance decaying to pristine photons.  We might therefore see ATLAS's preference for a larger width as a good sign for this type of model, despite the relatively narrow intrinsic width for $s$.

The strongest tension in this model arises
from $h\to a a$ decay.  This is particularly true if the desired cross-section is in the 5--10 fb range, where Br$(h\to aa)$ approaches 10\%.  Since the production of $s$ proceeds through mixing
with the Higgs, these decays cannot be suppressed arbitrarily.
However, there is an
alternate strategy for producing $s$ pairs without mixing with the
Higgs. If there are new colored
fermions which couple to the complex scalar $S$, but do \emph{not}
generate a mixed anomaly with QCD, the axion does not couple to
gluons and therefore still decays dominantly to a pair of photons. The
scalar $s$ does couple to gluons 
and
can be produced through gluon fusion. (The width of the state can also become tens of GeV by decreasing $f$, without causing unacceptable Higgs decays.) An amusing feature of this
model is that the production of $s$ is through the QCD scale anomaly, and the
decay of $a$ is through the axial anomaly.
In this class of models, $\sigma(pp\to s)$ can easily be in the 10 fb
range, $s$ mixing with the Higgs can be small so that Br$(h\to aa)$ is
small and Br$(s\to aa)\sim 1$, Br$(a\to \gamma\gamma)\sim 1$, and for
$m_a\sim 0.5-5$ GeV, the $a$ lifetime can be dialed to any value in
the interesting range of 0.1 -- 10 cm.\footnoteB{Note added: we will
give more details on these models in a future brief publication.}

We should also point out another very important signal that might
arise in a model of this type; it might be present in existing data,
and might have very low background.  Suppose the lifetime of $a$ is
long enough that occasionally an $a\to \gamma\gamma$ decay from $s\to
aa$ occurs beyond the ECAL and inside the hadronic calorimeter (HCAL).  Then a very rare
event would ensue: a ``photon'' from one $a$ decay,
with $p_T$ of several hundred GeV,
would recoil against a narrow, isolated, trackless, HCAL-only jet of
several hundred GeV from the other $a$ decay.  
If such
events are observed, and the HCAL energy
deposition is corrected for the fact that its origin was
electromagnetic, a peak would be observed at $m_s=750 \mathrm{\ GeV}$.
Relaxing the criteria for the ``photon'' (as in \cite{ATLAS:2012soa})
might further increase the observed signal.\footnoteB{Note added: this signal is unlikely to be observable in the near term within the model of Eq.~(\ref{eq:full-lag}), but easily arises in the models mentioned in the previous paragraph.}

Such HCAL-only jets have been used as
trigger-objects \cite{Aad:2013txa}, following
\cite{Strassler:2006im,Strassler:2006ri}, and as analysis objects
\cite{Aad:2014yea,Aad:2015asa}, following \cite{Falkowski:2010cm}.  In
\cite{Aad:2014yea} a search was carried out for objects decaying to
$e^+e^-$ pairs in the HCAL.   However, two such objects were required,
which is very rare for the lifetimes considered here, and there was
considerable background.  The search we propose for a hard photon plus
a hard HCAL-only jet, with no intrinsic missing energy in the event
once the HCAL energy is corrected, should have higher signal and much
lower background, especially since the parent particle's mass is
known.

\subsection{Dark photons looking like light}

An alternative scenario with some similar features would involve a
scalar or pseudoscalar $X_s$ coupled, as in (\ref{modelA}), to gluons,
photons, and hidden vector bosons $V_D$, which we will for simplicity
call ``dark photons''.  In particular the $X_sV_DV_D$ coupling can be
induced by particles in the hidden sector which could be related to
dark matter.  In this case we would have $gg\to X_s$ followed by
$X_s\to V_DV_D$.  We now  imagine the dark photons have mass $m_V<2
m_\mu$ and necessarily decay only to $e^+e^-$ via kinetic mixing with
the photon \cite{Holdom:1985ag,Jaeckel:2013ija,Essig:2013lka}. Such dark photons are inevitably displaced, given current
experimental constraints \cite{Jaeckel:2013ija,Essig:2013lka}.

By choosing SM and hidden matter appropriately, and choosing the dark
boson's gauge coupling, it is easy to get a cross-section times
branching fraction in the few fb range.   For instance, by taking
$\kappa_G$ smaller than in our earlier discussion and taking the
hidden coupling $\kappa_V$ greater than $\kappa_G,\kappa_\gamma$, we
can get $\sigma(pp\to X_s)\sim$ 5--10 fb and $\Br(X\to V_DV_D)\sim 1$.
In contrast to the model in Eq.~\eqref{eq:full-lag}, here $h\to
V_DV_D$ could be very small because we do not need to appeal to mixing
of $X_s$ and $h$ to produce the former.

With kinetic mixing such that the dark photon proper lifetime is, say, in
the 100 $\mu$m range, well within the allowed region
\cite{Pospelov:2008zw},
the boost of the
$V_D$ would be of several thousand and the lifetime in the lab frame
would be in the range of tens of cm.  
When both $V_D$ decay within the tracker, each would
appear to be a converted photon; therefore the peak would be more visible
in categories of events with one or two ``converted photons''.

Of course, other signs that this signal is not from photons would be present.  Mass resolution would again be somewhat worse, and the resonance therefore wider, than for a decay to two pristine photons.  The locations of apparent photon conversions in this signal would not correspond to the distribution of detector material but rather to a long-lived particle decay distribution.  And the (converted) photon plus HCAL-only jet  signal mentioned above, for the $s\to aa$ model, would potentially apply here as well.

\section{Quirks}
\label{sec:quirks}

The $X_s$ excess might also be interpreted as the ground state (or
possibly an overlay of several states) of a pair of quirks $Q$.
Quirks are particles that carry SM quantum numbers and hidden quantum
numbers, and are bound by a flux tube that, unlike those in QCD,
cannot easily break.\footnoteB{Particles of this type have been considered for
  decades \cite{Okun:1980kw}; the modern study of these objects dates
  to \cite{Kang:2008ea}, who gave them their modern name, and to
  \cite{Strassler:2006im}.  Their
    phenomenology has been
    considered in various contexts 
    in~\cite{Cai:2008au,Kang:2008ea,Burdman:2008ek,Harnik:2008ax,Juknevich:2009ji,Juknevich:2009gg,Harnik:2011mv,Burdman:2014zta,Curtin:2015fna}.} This type of flux tube
  arises in gauge theories where $Q$ is charged under a non-abelian
  gauge group with no light matter.  In this case $Q\bar Q$ can be
  created perturbatively, with a perturbatively calculable
  cross-section; but the $Q$ and $\bar Q$ cannot escape each other,
  and so they form a highly-excited bound state that deexcites somehow
  and eventually annihilates.  Both de-excitation and annihilation can
  give rise to glueballs and/or SM gauge bosons.  Here we will be
  considering hidden confinement scales near or above the QCD scale,
  and production that is not too far above threshold, so that the
  de-excitation and annihilation are prompt and the quirks do not
  travel macroscopic distances.

Quirks of mass at or near $375$ GeV would produce a diphoton resonance near 750 GeV.  
The quirks can be produced with any energy above 750 GeV, but they
then are trapped in a highly-excited bound state.  They de-excite by
radiating their excess energy as jets (often soft), photons, or hidden
hadrons (as we will discuss later.)  Having reached or at least
approached the ground state, with an invariant mass of or slightly
above 750 GeV, they then annihilate into hard SM or hidden gauge
bosons.\footnoteB{For some phenomenological study of quirks in this or
  nearby regimes, see \cite{Burdman:2008ek,Juknevich:2009ji,Juknevich:2009gg,Harnik:2011mv,Burdman:2014zta,Curtin:2015fna}.}  We would then identify $X_s$ as the ground state (or an overlay of a ground state with some nearby excited states) of quirkonium. 

Colorless quirks are not a very plausible explanation for the signal;
since they are produced in $q\bar q$ collisions, it is improbable that they would
have gone unseen at Run 1 but shown up at Run 2.  Suppose instead that
the quirks are color triplets.  For fermionic (scalar) quirks, the
cross-section is of order $N_h$ (the number of hidden colors) times
4.5 pb (0.5 pb) at 8 TeV, and 18 pb (2.6 pb) at 13 TeV.\footnoteB{The
  fermionic quirk production cross section is estimated by multiplying
  the LO result by a factor of 1.5. The scalar quirk production cross
  section is taken from NLO calculation of stop pair production cross
  section~\cite{Borschensky:2014cia}.}

For fermionic quirks, there is a significant obstacle. The low-lying spectrum (for both color-singlet and color-octet states) includes $0^{-}$ and $1^{-}$ states, analogous to the $\eta_c$ and $J/\psi$.  The spin-splittings will be relatively small, as in known quarkonium systems, and the spin-singlet and spin-triplet states will be populated at similar rates.  The spin singlet will have $\Br(0^+\to jj)/\Br(0^+\to \gamma\gamma) \sim 45/q^4$, where $q$ is the quirk's electric charge.
But the color-singlet spin-triplet state will create a dilepton resonance at a higher production rate than the diphoton resonance from the $0^-$ state.  This is similar to what happens for the $\eta_c$ and $J/\Psi$; unlike the $0^-$ state, the $1^-$ state can only decay to three hidden or SM gluons, and so its decay via a single hypercharge boson is important.  We therefore expect the rate for $1^-\to\ell^+\ell^-$ to exceed $0^-\to \gamma\gamma$ substantially. Since 8 TeV limits on 750 GeV dilepton resonances are in the 1 fb range, it is hard to believe this would not have been observed.  Because of the uncertainties inherent in quirk calculations, there might be a loophole, but we view this option as disfavored.  Still, dilepton resonances just above 750 GeV might well appear.

Scalar quirks potentially have a naturalness problem and therefore are somewhat less appealing, but it is worth pointing out that they are better phenomenologically.  First, the cross-section is not as dangerously large as for fermionic quirks.  Second, the ground state S-wave is just a $0^+$, with no corresponding $1^-$ state; the lowest $1^-$ state is in the P-wave.  This state is therefore split further in mass from the $0^-$ than in the fermionic case, and its annihilation rate is much lower, with the effect that a radiative decay is more likely than annihilation.  It seems likely, therefore, that the dilepton resonance is removed for scalar quirks.

However, that is not the end of the story.
We must consider quirkonium states that are octets of $SU(3)_c$ as well as singlets.   Octet states can decay neither to hidden gluons nor to electroweak bosons, so they must either radiate a gluon or annihilate to jets or to jet + electroweak boson. Depending on the pattern of de-excitation, the probability of ending up in the $0^+$ octet states may be up to $\sim 8$ times more likely than ending up in the $0^+$ singlet state.  This dilutes, by a factor $x$ which could be as small as  $1/9$, the cross-section for the $0^+$ state and correspondingly the di-photon signal.  However, because quirks in the octet have a repulsive QCD interaction, they may be split from the singlet states, the amount depending on the hidden sector coupling $\alpha_h$, and this may bring $x$ back toward 1.

The fraction of color-singlet $0^+$ states decaying to hidden gluons, SM gluons and SM photons is in the ratio (in the  perturbative regime, and at leading order)
\beq \label{dkratios}
\left (C_h
\alpha_h \right)^2 : \left(\frac{4}{3}\alpha_s\right)^2 : 8 Q^4 \alpha^2 .
\eeq
where $C_h\equiv (N_h^2-1)/2N_h$.
This estimate comes with subtleties, but let us first assume
$\alpha_h=\alpha_s$ and $N_h=3$ to see what happens. Then the 13 TeV
cross-section is about $7.8$ pb (and $1.5$ pb at 8 TeV) for the scalar
quirks.\footnoteB{This cross-section has a large uncertainty.
Hidden-QCD corrections will increase it.  Meanwhile the binding energy
of the $0^+$ state implies that the scalar quirk mass is not half the
resonance mass.}  Since the branching fraction to hidden gluons and SM
gluons are then equal, $3.4x$ pb of $0^+$ states decay to gluons, an
equal amount decay to hidden gluons and from there to hidden glueballs
(about which we will say more below), and $67 x q^4$ fb decay to
photons.  We get an acceptable signal for $q=2/3, x\sim 1/3$, 
within the uncertainties.

A number of things could change this estimate.  If $\alpha_h$ is significantly smaller than $\alpha_s$, decays to hidden glueballs become rather rare and the cross-section for photons nearly doubles, which is acceptable with a smaller $x$, or a smaller $q$ and larger $x$.

If $\alpha_h$ and/or $N_h$ is much larger, the estimate quickly breaks down.  Indeed if $\alpha_h N_h > .9$, hidden confinement occurs at $>50$ GeV.  This makes the hidden glueballs heavier than $375$ GeV, forbidding squirks to decay to them.  In this case the hidden decays shut off.  For values somewhat less, the diphoton rate becomes incalculable.  

If octet states commonly decay to dijets and/or singlet states rarely
decay to hidden glueballs, then they may combine to form a dijet
resonance.  Run 1 limits on a 750 GeV dijet resonance are at about 2
pb (for $gg$ initial states) \cite{CMSdijet} which is quite close to
our squirk cross-section for $N_h=3$, so a dijet resonance may be
visible at Run 2 near 750 GeV --- assuming data scouting methods can
be applied down to such low  dijet mass, which is not necessarily the
case. The resonance may have an odd shape if the splitting of octet
and singlet states is sufficiently large.

Also interesting to consider is that octet states can decay to photon + jet, so a resonance would be expected there as well if $x$ is small.  This has the advantage that triggering will not be a problem, but the disadvantage that the signal's size depends on parameters and on the unknown dilution factor $x$.

Although annihilation is most common after deexcitation is over, it can occur during deexcitation, as seen in the excited charmonium and bottomonium states.  As a result, several photon peaks, possibly overlapping or merged into a single misshapen peak, are possible in these models.  To the extent the states are vaguely hydrogenic, the splitting between states is of order
\beq
\Delta m_{n_1,n_2} \sim
\frac{1}{4}
C_h^2 \alpha_h^2 m
\left(\frac{1}{n_1^2}-\frac{1}{n_2^2}\right)
\eeq
where $n_i$ are initial and final radial quantum numbers, and
$\alpha_h$ is the hidden coupling evaluated at the radius of the bound
state.  Such an estimate is only accurate at order 1 but shows the
splitting between the ground and first excited state could easily be
in the range of a few GeV to a few tens of GeV. If $\alpha_h$ is large
then the linear potential between the quirks can increase the
splitting further.

A further phenomenon expected for quirks is the particles produced
during de-excitation.  For colored quirks these may be soft jets or
even stray pions.  They may also include light hidden glueballs, whose
final states are model dependent but which are often long-lived.
Signs of these phenomena could therefore include (1) a large number of
tracks at the primary vertex compared to the background, often
carrying energy of order 100 - 200 GeV; (2) small amounts of missing
transverse momentum, usually less than 100 - 200 GeV; (3) long-lived
glueballs (which might cause some of the diphoton events to be
discarded.)  It would be interesting to compare the events within the
diphoton peak with those outside it with respect to these measures.

From (\ref{dkratios}) the rate for production of hidden glueballs would be, in the perturbative regime,
\beq
N_h\times ( 2.6 \ {\rm pb} )
\frac{C_h \alpha_h^2}{ \frac{4}{3}\alpha_s^2+C_h\alpha_h^2} .
 \eeq
This could be several pb, giving tens of thousands of events at Run 1
and thousands already at Run 2.  However, as we noted, this rate could
also be zero non-perturbatively.  The phenomenology of the glueballs
is potentially quite interesting.  The lightest CP-even glueballs of a
hidden sector, if coupled to the SM only through the quirks, will
decay to SM dibosons of all types \cite{Juknevich:2009ji},
with branching fraction ratios similar to those for the
$0^+$ quirkonium decay to SM dibosons.   They can therefore give lower
mass diphoton resonances which may currently lie buried in background.
(Rare four-photon events would also appear.) The glueball lifetimes
vary as their mass to the 9th power, and so sweep over many orders of
magnitude.  With so many regimes available, we leave this for future
consideration.

Quarkonium states have a moderate width; in QCD the $\eta_c$ has a
width of 1\% of its mass, which here would be $\sim 8$ GeV.  
Also, as noted earlier, the excited $0^+$ states
might perhaps generate additional diphoton peaks and distort the apparent
resonance, making it appear initially as a single, wider resonance.

In summary, fermionic colored quirks seem unlikely because a dilepton
resonance would be expected to have been observed, but scalar colored
quirks seem more plausible.  Dijet and jet + photon resonances at or
near (and above) 750 GeV are expected but do not appear to be
excluded.  
Further signatures may include additional weaker diphoton peaks above the main one,
separated by a few GeV or more, which might blend into a single peak
with a tail at higher mass.   All of these events may be accompanied
by unusual track activity, small but noticeable missing energy, or long-lived
particles.  There might also be additional weak diphoton peaks at
much lower mass from glueball decays, which would appear in events with two
or more jets.

\section{Conclusions} \label{sec:conclusions}

In this paper we considered several classes of models that could lie behind the excess in the ATLAS and CMS diphoton searches.  Our goal was to suggest avenues of investigation for the experimental collaborations that might allow a signal to be identified now or soon, before the significance of the diphoton bump could reach discovery significance on its own.  

The first possibility is the production of a pseudoscalar which decays
to two photons through loops of heavy colored and charged vector-like
quarks. Though it is a trivial explanation, interesting new signals
arise when one tries to embed it in a natural scenario such as little
Higgs with a pseudoscalar and an approximate T-parity. The vector-like
quarks, which are top or bottom partners in a natural scenario, might
decay dominantly to the pseudoscalar plus top or bottom. In this case,
pair production of multiple $\sim$ TeV mass top partners, followed by
$t'\to t$ plus the new resonance,
would give $t\bar t$ plus four jet events.  These
could be searched for by reinterpreting 13 TeV SUSY searches with many jets and low missing energy and/or a lepton. 

A distinctively different situation is that the ``photons" in the excess are actually long-lived particles decaying to photon pairs or to electron pairs. This could be realized in simple model with a complex scalar, in which the real 750 GeV scalar decays to two light boosted pseudoscalars and the pseudoscalar decays to two collimated photon pairs subsequently. 
 In this case, there are opportunities for detecting overlapping photons and/or unusual patterns of apparent photon-conversions in either the 750 GeV scalar or 125 GeV Higgs decays. There is also the possibility of events with a hard ``photon'' recoiling against a narrow isolated HCAL-only jet; the photon-jet invariant mass, once the jet's energy is corrected for its electromagnetic source, would show a peak at 750 GeV.  We also considered a similar model with dark photons decaying late to $e^+e^-$ pairs.  It is possible that differences in the ATLAS and CMS detectors, and their definitions of ``photon'' in different analyses, could lead to the differences in observed excesses.
 
 Finally the 750 GeV bump could be from quirkonium, a bound state of quirks.  These particles are permanently bound by a confining force after production, and the de-excitation and annihilation of their bound states could give rise to photon pairs.  These events would often be accompanied by a large number of soft tracks or soft jets, and perhaps small missing energy or even long-lived particles. Furthermore, there could be dijet and jet + photon resonances at or near (and above) 750 GeV, and perhaps  a dilepton resonance.   There might also be signatures of hidden glueball decays such as additional weak diphoton peaks at lower mass, seen in events with at least two jets.

Simple models of the type we have mentioned above tend to give
relatively narrow resonances, though cases with false photons will
have worse resolution, widening the state.  However, as we noted, the
widths can be increased with mild changes in the models, so if the
$X_s$ proves to be both real and wide, the models considered
here can accommodate it.  Importantly, the signals of new physics that we have described here need not be lost in such a case.

We hope our experimental colleagues will search for these  signals in the near future, especially since the analyses we propose are interesting and sensitive to new phenomena of various sorts, even if the $X_s$ itself proves to be an merely an excess born of low statistics.

\section*{Acknowledgments}

The work of MR is supported in part by the NSF Grant PHY-1415548.
The work of PA was supported in part by NSF grants
PHY-0855591 and PHY-1216270. BH is supported by the Fundamental Laws
Initiative of the Center for the Fundamental Laws of Nature. 

\bibliography{ref}
\bibliographystyle{utphys}
\end{document}